\documentclass[aip, superscriptaddress]{revtex4-1}

\pdfoutput=1
\usepackage{graphicx}

\usepackage{amsmath}
\usepackage{amssymb}
\usepackage{bm}
\usepackage{ulem}

\newcommand{\mean}[1]{\langle{#1}\rangle}

\newcommand{\bra}[1]{\langle{#1}|}
\newcommand{\ket}[1]{|{#1}\rangle}

\newcommand{\Tr}{{\rm Tr}\hspace{0.07cm}}

\newcommand{\abs}[1]{{|#1|}}

\begin{document}

\title{
Quantum asymptotic amplitude for quantum oscillatory systems from the Koopman operator viewpoint
}

\author{Yuzuru Kato}
\email{Corresponding author: katoyuzu@fun.ac.jp}
\affiliation{Department of Complex and Intelligent Systems, Future University Hakodate, Hokkaido 041-8655, Japan }

\date{\today}

\begin{abstract}
	We have recently proposed a fully quantum-mechanical definition of the asymptotic phase for quantum nonlinear oscillators, which is also applicable in the strong quantum regime [Kato and Nakao 2022 Chaos 32 063133]. In this study, we propose a definition of the quantum asymptotic amplitude for quantum oscillatory systems, which naturally extends the asymptotic amplitude for classical nonlinear oscillators on the basis of the Koopman operator theory. We introduce the asymptotic amplitude for quantum oscillatory systems by using the eigenoperator of the backward Liouville operator associated with the largest non-zero real eigenvalue. Using examples of the quantum van der Pol oscillator with the quantum Kerr effect, which exhibits quantum limit-cycle oscillations, and the quantum van der Pol model with the quantum squeezing and degenerate parametric oscillator with nonlinear damping, which exhibit quantum noise-induced oscillations,  we demonstrate that the proposed quantum asymptotic amplitude yields isostable amplitude values that decay exponentially with a constant rate. Furthermore, using the quantum asymptotic amplitude, we introduce effective quantum periodic orbits for quantum limit-cycle oscillations and quantum noise-induced oscillations. 
\end{abstract}

\maketitle

\section{Introduction}

Rhythmic oscillations are ubiquitous in nature and have been extensively studied in a wide range of classical systems, including heartbeats, circadian rhythms, chemical oscillations, and electrical oscillators  \cite{winfree2001geometry, kuramoto1984chemical, pikovsky2001synchronization, nakao2016phase, ermentrout2010mathematical, strogatz1994nonlinear}.
With recent advances in quantum technology,  a theoretical model of quantum rhythmic oscillations \cite{lee2013quantum, walter2014quantum} has been proposed and  their novel features  have been extensively studied
 \cite{
 	lee2013quantum,
 	walter2014quantum, 
 	sonar2018squeezing,
 	lee2014entanglement,
 	kato2019semiclassical,
 	lorch2016genuine,
 	mok2020synchronization, 
 	witthaut2017classical,
 	roulet2018quantum,
 	lorch2017quantum,
 	weiss2016noise,
 	es2020synchronization,
 	kato2021enhancement, 
 	li2021quantum, 
 	kato2020semiclassical,
 	hush2015spin,
 	weiss2017quantum,
 	mari2013measures,
 	xu2014synchronization,
 	roulet2018synchronizing,
 	chia2020relaxation, 
 	arosh2021quantum, 
 	jaseem2020generalized, 
 	jaseem2020quantum, 
 	cabot2021metastable, 
 	kato2022definition,
 	kato2023quantum, setoyama2024lie, bandyopadhyay2023aging}. 
 In addition, experimental observations of phase synchronization between quantum rhythmic oscillations have also been reported \cite{laskar2020observation, krithika2022observation, koppenhofer2020quantum}.

Rhythmic oscillations in classical deterministic systems can be modeled as nonlinear dynamical systems with stable limit-cycle solutions. The concepts of \textit{asymptotic phase} and \textit{isochron} \cite{winfree2001geometry, kuramoto1984chemical, pikovsky2001synchronization, nakao2016phase, ermentrout2010mathematical}, which exhibits  a linear increase with a constant frequency within the basin of attraction of the limit cycle, are commonly used for analyzing synchronization phenomena. 
The asymptotic phase, originally introduced  by Winfree \cite{winfree1967biological} and Guckenheimer \cite{guckenheimer1975isochrons} from a geometrical perspective  \cite{winfree1967biological, guckenheimer1975isochrons}, has been shown to be closely related to the Koopman eigenfunction associated with the fundamental oscillation frequency \cite{mauroy2013isostables, shirasaka2017phase, mauroy2020koopman, kuramoto2019concept, mauroy2018global, shirasaka2020phase}.

From the Koopman perspective, the concepts of \textit{asymptotic amplitudes} and \textit{isostables}, which characterize deviation from the limit cycle and exhibit exponential decay with constant rates toward the limit cycle, have been naturally introduced in terms of the Koopman eigenfunction associated with the Floquet exponents having non-zero real parts \cite{mauroy2013isostables, shirasaka2017phase, mauroy2020koopman, kuramoto2019concept, mauroy2018global, kvalheim2021existence, shirasaka2020phase}. 
These phase and amplitude functions are fundamental quantities for characterizing limit-cycle oscillations.  Moreover,  when the system dynamics are dominated by slow modes \cite{shirasaka2017phase}, a common feature in realistic oscillator models 
\cite{monga2019phase}, a phase-amplitude reduction can be applied \cite{
	mauroy2013isostables, shirasaka2017phase, mauroy2020koopman, kuramoto2019concept, mauroy2018global, wilson2016isostable,  monga2019phase, shirasaka2020phase}. This reduction technique provides a useful framework for analyzing and controlling synchronization dynamics.

The phase and amplitude description can also be extended to classical stochastic oscillatory systems. The stochastic asymptotic phase \cite{thomas2014asymptotic}, which exhibits a linear increase with a constant frequency on average under the system's stochastic dynamics, has been introduced using the slowest-decaying eigenfunction of the backward Fokker-Planck (Kolmogorov) operator. Similarly, the stochastic asymptotic amplitude \cite{perez2021isostables}, which decays exponentially with a constant rate on average under the system's stochastic dynamics,  has been introduced based on the eigenfunction associated with the largest non-zero real eigenvalue. These definitions of stochastic asymptotic phase and amplitude are closely related to stochastic Koopman operator theory \cite{junge2004uncertainty, vcrnjaric2020koopman, wanner2020robust} and
can be viewed as natural extensions of their deterministic counterparts defined based on Koopman eigenfunctions \cite{kato2021asymptotic}.
Moreover, the zero-level set of the stochastic asymptotic amplitude defines a periodic orbit associated with the stochastic oscillatory dynamics \cite{perez2021isostables}. When the deterministic part of the system possesses a stable limit cycle, this zero-level set coincides with the limit-cycle attractor  in the noiseless limit.

By extending the definition of the stochastic asymptotic phase from the Koopman operator viewpoint, we recently introduced  a definition of \textit{the quantum asymptotic phase} for quantum oscillatory systems. This quantum asymptotic phase exhibits linear increase under the  system's quantum dynamics and provides isochronous phase values for analyzing quantum synchronization even in the strong quantum regime \cite{kato2022definition,kato2023quantum}. However, the concept of \textit{quantum asymptotic amplitude} for quantum oscillatory systems remains unexplored, and quantum periodic orbits associated with quantum oscillatory dynamics are not yet well characterized.

In this study, we introduce a definition of the quantum asymptotic amplitude for quantum oscillatory systems by extending the concept of the stochastic asymptotic amplitude from the Koopman operator viewpoint. Specifically, the quantum asymptotic amplitude is defined in terms of the eigenoperator of the backward Liouville operator associated with the largest non-zero real eigenvalue. Using examples such as the quantum van der Pol oscillator with the quantum Kerr effect, which exhibits quantum limit-cycle oscillations, and the quantum van der Pol model with quantum squeezing, as well as the degenerate parametric oscillator with nonlinear damping, which exhibit quantum noise-induced oscillations, we numerically demonstrate that the proposed quantum asymptotic amplitude yields appropriate amplitude values that decay exponentially at a constant rate under quantum system dynamics.
Furthermore, based on the quantum asymptotic amplitude, we introduce effective quantum periodic orbits for both quantum limit-cycle oscillations and noise-induced oscillations, which 
characterize the quantum oscillatory behavior that inherently reflects quantum effects.

\section {Asymptotic ampliutde for quantum oscillatory systems}
\label{sec:quantum}

In this section, we provide a brief overview of quantum oscillatory systems, review the quantum asymptotic phase introduced in Ref.~\cite{kato2022definition}, and present a definition of the quantum asymptotic amplitude. 

\subsection{Quantum oscillatory systems} 
\label{sec:quantum_me}

We consider an open quantum system that exhibits oscillatory behavior.
Assuming that the interactions between the system and its reservoirs occur instantaneously, we adopt the Markovian approximation. Under this assumption, the time evolution of the system's density operator  $\rho$ is governed by the quantum master equation~\cite{carmichael2007statistical, gardiner1991quantum,breuer2002theory}
\begin{align}
\dot{\rho}
= \mathcal{L}\rho =
-i[H, \rho] 
+ \sum_{j=1}^{n} \mathcal{D}[C_{j}]\rho,
\label{eq:me}
\end{align}
where $( \dot{} )$ denotes the time derivative, $\mathcal{L}$ denotes a Liouville superoperator governing
the evolution of $\rho$,
$H$ denotes a system Hamiltonian, 
$C_{j}$ denotes a coupling operator between the system  and the
$j$th reservoir $(j=1,\ldots,n)$,  
$[A,B] = AB-BA$ denotes the commutator,
$\mathcal{D}[C]\rho = C \rho C^{\dag} - (\rho C^{\dag} C + C^{\dag} C \rho)/2$ 
denotes the Lindblad form where $\dag$ indicates the Hermitian conjugate, 
and we set the reduced Planck's constant as $\hbar = 1$.

We introduce an inner product $\mean{X,Y}_{tr} =\Tr{ ( X^{\dag}Y )}$ of linear operators $X$ and $Y$ and define the superoperator $\mathcal{L}^{*}$, the adjoint of $\mathcal{L}$, such that it satisfies
$\mean{\mathcal{L}^{*}X,Y}_{tr}
=\mean{X,\mathcal{L}Y}_{tr}$, which can be explicitly expressed as 
\begin{align}
\mathcal{L}^{*} X = i[H, X] + \sum_{j=1}^{n} \mathcal{D}^{*}[C_{j}]X,
\end{align}
where $\mathcal{D}^{*}[C]X = C^{\dag} X C - (X C^{\dag} C + C^{\dag} C X)/2$ is the adjoint Lindblad form. 

The superoperator $\mathcal{L}^{*}$ governs the evolution of an observation operator $F$ through the equation
\begin{align}
\dot{F} = {\cal L}^{*} F.
\label{eq:bme}
\end{align}

In the Schr\"odinger picture, the density operator $\rho$ evolves according to Eq.~(\ref{eq:me}), whereas the observation operator $F$ remains constant.  Conversely,  in the Heisenberg picture,  $F$ evolves according to Eq.~(\ref{eq:bme}), whereas $\rho$ remains constant.
In both pictures, the expectation value of $F$ with respect to $\rho$, namely, 
\begin{align}
	\langle F \rangle_q = \mbox{Tr}(\rho F) = \mean{\rho, F}_{tr}
	\label{efpave_qt}
\end{align}
remains the same value ($\rho$ is self-adjoint).

The Liouville superoperator $\mathcal{L}$ possesses a biorthogonal eigensystem $\{\Lambda_{k}, U_{k}, V_{k}\}_{k=0, 1, 2, ...}$, which consists of the eigenvalue $\Lambda_{k}$ and eigenoperators $U_{k}$ and $V_{k}$, satisfying
\begin{align}
\mathcal{L} U_{k} =  \Lambda_{k} U_{k},
\quad
\mathcal{L}^{*} V_{k} =  \overline{\Lambda_{k}} V_{k},
\quad
\mean{V_{k}, U_{l}}_{tr} = \delta_{kl},  %\cr ~ (k, l=0, 1, 2, \ldots)
\label{eq:eigentriplet1}
\end{align}
for $k, l=0, 1, 2, \ldots$, where the overline denotes the complex conjugate. 

We assume that one eigenvalue, denoted $\Lambda_0$, equals zero, corresponding to the stationary state $\rho_0$ of the system, namely ${\cal L} \rho_0 = 0$, and that all other eigenvalues have negative real parts. This assumption also implies the absence of decoherence-free subspaces \cite{lidar1998decoherence}. 
Furthermore, we assume that the eigenvalues with the smallest absolute real part, that is, those associated with the slowest decay rate, form a complex-conjugate pair,
\begin{align}
\Lambda_1 = \mu_q - i \Omega_q,
\quad
\overline{ \Lambda_1 } = \mu_q + i \Omega_q,
\end{align}
where $\abs{\mu_q}~(\mu_q < 0)$ represents the decay rate, and $\Omega_q = \mbox{Im}\ \overline{ \Lambda_1 }$ represents the frequency of the fundamental oscillation. 
With these assumptions, the system exhibits oscillatory behavior.

\subsection{Phase-space representation}
\label{sec:quantum_qpro}

We can transform the density operator $\rho$ into a quasiprobability distribution in the phase space~\cite{carmichael2007statistical, gardiner1991quantum, cahill1969density}.
By employing the $P$-representation~\cite{carmichael2007statistical, gardiner1991quantum, cahill1969density},  the density operator $\rho$ 
is expressed as
\begin{align}
\rho = \int p({\bm \alpha}) | \alpha \rangle \langle \alpha |  d {\bm \alpha},
\label{eq:prep}
\end{align}
where $| \alpha \rangle$ represents a coherent state characterized by a complex value $\alpha$, or a complex vector $\bm{\alpha} = (\alpha, \overline{\alpha})^{T}$, the function $p({\bm \alpha})$ represents a quasiprobability distribution of ${\bm \alpha}$, $d{\bm \alpha} = d\alpha d\overline{\alpha}$, and the integral is taken over the whole complex plane.

Similarly, we transform the observable $F$ into a function in the phase space as
\begin{align}
f({\bm \alpha}) = \langle \alpha | F | \alpha \rangle,
\end{align}
where the operator $F$ is arranged in the normal order~\cite{carmichael2007statistical, gardiner1991quantum, cahill1969density}.
Introducing the $L^2$ inner product $\mean{g(\bm{\alpha}), h(\bm{\alpha})}_{\bm{\alpha}} = \int \overline{ g(\bm{\alpha}) } h(\bm{\alpha}) d \bm{\alpha}$ for two functions $g(\bm{\alpha})$ and $h(\bm{\alpha})$,  we express the expectation value of $F$ with respect to $\rho$ as
\begin{align}
\langle F \rangle_q = \mbox{Tr} (\rho F) = \int d{\bm \alpha} p({\bm \alpha}) f({\bm \alpha}) = \mean{p(\bm{\alpha}), f(\bm{\alpha})}_{\bm{\alpha}}.
\end{align}

We describe the time evolution of $p({\bm \alpha})$, which corresponds to Eq.~(\ref{eq:me}), by a partial differential equation
\begin{align}
\partial_t{p}({\bm \alpha}) = {L}_{\bm \alpha} p({\bm \alpha}),
\label{eq:qpde}
\end{align}
where the differential operator ${L}_{\bm \alpha}$ 
satisfies $\mathcal{L}\rho = \int {L}_{\bm \alpha} p({\bm \alpha}) | \alpha \rangle \langle \alpha | d{\bm \alpha}$.
We can explicitly derive ${L}_{\bm \alpha}$ from Eq.~(\ref{eq:me}) using the correspondence of the quantum operator to the differential operator in the phase-space representation \cite{carmichael2007statistical, gardiner1991quantum}. 
In the Heisenberg picture, correspondingly, the evolution of $f({\bm \alpha})$  is similarly described by
\begin{align}
\partial_t f({\bm \alpha}) = {L}^{+}_{\bm \alpha}f({\bm \alpha}),
\end{align}
where the differential operator ${L}^{+}_{\bm \alpha}$ denotes
the adjoint of ${L}_{\bm \alpha}$
with respect to the $L^2$ inner product, namely, 
$ \mean{ {L}^{+}_{\bm \alpha}g(\bm{\alpha}), h(\bm{\alpha})}_{\bm{\alpha}}
=\mean {g(\bm{\alpha}), {L}_{\bm \alpha} h(\bm{\alpha})}_{\bm{\alpha}}$,
which satisfies ${L}^{+}_{\bm \alpha}f({\bm \alpha}) = 
\langle \alpha | \mathcal{L}^{*} F | \alpha \rangle$.

The differential operator ${L}_{\bm \alpha}$ possesses a biorthogonal eigensystem $\{\Lambda_{k}, U_{k}, V_{k}\}_{k=0, 1, 2, ...}$, which
consists of the eigenvalue $\Lambda_{k}$ and eigenfunctions ${u}_{k}({\bm \alpha})$ and  ${v}_{k}({\bm \alpha})$, satisfying
\begin{align}
&{L}_{\bm \alpha} {u}_{k} = \Lambda_{k} {u}_{k},
\quad
{L}^{+}_{\bm \alpha}{v}_{k} = \overline{\Lambda_{k}} {v}_{k},
\quad
\mean{ {v}_{k}, {u}_{l}}_{\bm \alpha} 
= \delta_{kl}.
\end{align}

This eigensystem corresponds one-to-one with Eq.~(\ref{eq:eigentriplet1}).
The eigenvalues $\{ \Lambda_{k} \}_{k \geq 0}$ are identical to those of ${\cal L}$. The eigenfunctions ${u}_{k}$ and ${v}_{k}$ of ${L}_{\bm \alpha}$ and ${L}^{+}_{\bm \alpha}$ are linked to the eigenoperators $U_{k}$ and $V_{k}$ of $\mathcal{L}$ and $\mathcal{L}^{*} $, respectively, as
\begin{align}
U_{k} = 
\int {u}_{k}({\bm \alpha}) 
| \alpha \rangle \langle \alpha |
d{\bm \alpha},
\quad
{v}_{k}({\bm \alpha}) = 
\langle \alpha | V_{k} | \alpha \rangle,
\end{align}
which are obtained from 
$\mathcal{L} U_{k}
= 
\int {u}_{k}({\bm \alpha}) \left\{ \mathcal{L} 
| \alpha \rangle \langle \alpha | \right\} d{\bm \alpha}
= 
\int \left\{ {L}_{\bm \alpha} {u}_{k}({\bm \alpha}) \right\}
| \alpha \rangle \langle \alpha | d{\bm \alpha}
=
\int \Lambda_k {u}_{k}({\bm \alpha}) 
| \alpha \rangle \langle \alpha | d{\bm \alpha}
=
\Lambda_{k} U_{k}
$
and
$
{L}^{+}_{\bm \alpha}{v}_{k} = {L}^{+}_{\bm \alpha}\langle \alpha | V_{k} | \alpha \rangle
= 
\langle \alpha | {\mathcal L}^* V_{k} | \alpha \rangle = \overline{\Lambda_{k}} \langle \alpha | V_{k} | \alpha \rangle
=
\overline{\Lambda_{k}} {v}_{k}.
$

\subsection{Quantum asymptotic phase}
\label{sec:quantum_qphs}

Before introducing the quantum asymptotic amplitude, we briefly review  the quantum asymptotic phase
as introduced in Ref.~\cite{kato2022definition}.

In Ref.~\cite{kato2022definition}, we introduced the quantum asymptotic phase $\Phi_q(\rho)$ of a quantum state $\rho$ as the argument of the expectation value of the eigenoperator ${V}_1$ associated with the 
eigenvalue $\overline{\Lambda_1}$ as 
\begin{align}
	\Phi_q(\rho) = \arg \langle V_1 \rangle_q = \arg \langle \rho, V_1 \rangle_{tr} = \arg \langle p({\bm \alpha}), v_1({\bm \alpha}) \rangle_{\bm \alpha}.
\end{align}
This quantum asymptotic phase increases with a constant frequency $\Omega_q$ when the quantum state evolves according to Eq.(\ref{eq:me}), namely, 
\begin{align}
	\frac{d}{dt} {\Phi}_q(\rho) = \Omega_q.
\end{align}

In particular, the asymptotic phase of the coherent state $\rho^{\alpha} = \ket{\alpha}\bra{\alpha}$ 
can be regarded as a quantum asymptotic phase $\Phi_q({\bm \alpha})$ of the coherent state $\bm{\bm \alpha}$ in the $P$ representation, expressed as 
\begin{align}
	\Phi_q({\bm \alpha}) = \arg \langle \rho^{\alpha}, V_1 \rangle_{tr} = \arg { {v}_1({\bm \alpha}) } = \arg \langle \alpha | { V_1 } | \alpha \rangle
	\label{eq:phs_qt}.
\end{align}

The quantum asymptotic phase naturally extends from the stochastic asymptotic phase, which is introduced in terms of the slowest decaying eigenfunction of the backward Kolmogorov (Fokker-Planck) operator \cite{thomas2014asymptotic} and coincides with the deterministic asymptotic phase in the noiseless limit from the Koopman operator viewpoint \cite{kato2021asymptotic}.
Indeed,  when the quantum system in the classical limit possesses a stable-limit cycle attractor, the quantum asymptotic phase coincides with the deterministic asymptotic phase in the classical limit \cite{kato2022definition}.

\subsection{Quantum asymptotic  amplitude}
\label{sec:quantum_qiso}

The aim of this study is to introduce a definition of the quantum asymptotic amplitude by extending the concept of the stochastic asymptotic amplitude \cite{perez2021isostables}, in a manner analogous to the definition of the quantum asymptotic phase.  

For the asymptotic amplitude, we select the largest non-zero real eigenvalue, denoted by $\kappa_q = \Lambda_2$, considering it characterizes the decay rate of the asymptotic amplitude of the system (see also Fig.~\ref{fig_1}(a) and Fig.~\ref{fig_2}).

We define the quantum asymptotic amplitude $R_q(\rho)$ of the quantum state $\rho$ as the expectation value of the eigenoperator ${V}_2$, which is associated with the eigenvalue $\kappa_q (= \Lambda_2)$, as 
\begin{align}
	R_q(\rho) 
	= \langle  p({\bm \alpha}), { v_2({\bm \alpha}) } \rangle_{\bm \alpha} 
	= \langle \rho, { V_2} \rangle_{tr}.
\end{align}

We can confirm that the asymptotic amplitude $R_q(\rho)$ evolves with a constant rate $\kappa_q$ under the time evolution of the quantum state $\rho$ as governed by  Eq.(\ref{eq:me}), 
\begin{align}
	\frac{d}{dt} R_q(\rho) 
	& = \left \langle \frac{\partial p({\bm \alpha})}{\partial t}, \ { v_2({\bm \alpha}) } \right \rangle_{\bm \alpha}
	= \langle  L_{\bm \alpha} p({\bm \alpha}), \ { v_2({\bm \alpha}) } \rangle_{\bm \alpha} 
	=
	\langle  p({\bm \alpha}), \ {L}^{+}_{\bm \alpha} { v_2({\bm \alpha}) } \rangle_{\bm \alpha} 
	\cr
	&=
	\langle   p({\bm \alpha}), \kappa_q { v_2({\bm \alpha}) } \rangle_{\bm \alpha} 
	=
	\kappa_q \langle   p({\bm \alpha}), { v_2({\bm \alpha}) } \rangle_{\bm \alpha} 
	=
	\kappa_q R_q(\rho),
\end{align}
or equivalently,
\begin{align}
	%\hspace{-1em}
	\frac{d}{dt} R_q(\rho) 
	&= \langle \dot{\rho}, V_2 \rangle_{tr}
	= \langle {\mathcal L} {\rho}, V_2 \rangle_{tr}
	= \langle  \rho, {\mathcal L}^* V_2 \rangle_{tr}
	\cr
	&= \langle \rho,\kappa_q V_2 \rangle_{tr}
	=\kappa_q \langle  \rho, V_2 \rangle_{tr}
	=\kappa_q R_q(\rho).
\end{align}
Integrating over time, we obtain
\begin{align}
	R_q(\rho) = \exp( \kappa_q t ) R_q(\rho_0),
\end{align}
where $\rho_0$ is the initial state at $t=0$.
Thus, the asymptotic amplitude $R_q(\rho)$ decays exponentially with a constant rate $\kappa_q$ when the quantum state $\rho$ evolves according to Eq.~(\ref{eq:me}).

Analogous to the quantum asymptotic phase,  the asymptotic amplitude of the coherent state $\rho^{\alpha} = \ket{\alpha}\bra{\alpha}$ can be interpreted as a quantum asymptotic amplitude $R_q({\bm \alpha})$ of the coherent state $\bm{\bm \alpha}$ in the $P$ representation, expressed as
\begin{align}
	\label{eq:qiso}
	R_q({\bm \alpha}) =  \langle \rho^{\alpha}, V_2 \rangle_{tr} = {v}_2({\bm \alpha})  =  \langle \alpha | { V_2 } | \alpha \rangle.
\end{align}
The quantum asymptotic amplitude extends naturally from the asymptotic amplitude of the stochastic oscillators, which is introduced in terms of the eigenfunction of the backward Kolmogorov (Fokker-Planck) operator \cite{perez2021isostables} associated with the largest non-zero real eigenvalue, and coincides with the deterministic asymptotic amplitude in the noiseless limit from the Koopman operator viewpoint \cite{kato2021asymptotic}. 
%

%%%
Moreover,  we introduce the effective quantum periodic orbit $\chi_q$ as the zero-level set of the quantum asymptotic amplitude, namely,
%%%
\begin{equation}
\chi_q = \{ \bm{\alpha} | R_q(\bm \alpha) = 0 \},
\end{equation}
when this set forms  a periodic orbit. When the quantum system in the classical limit possesses a stable-limit cycle attractor, this effective quantum periodic orbit coincides with the deterministic limit-cycle trajectory in the classical limit. 
%%%

It is important to note that the quantum asymptotic phase $\Phi_q(\rho)$ and amplitude $R_q(\rho)$ of the quantum state $\rho$ can be directly derived from the quantum master equation without resorting to the phase space representation.  We employed the $P$ function to introduce the quantum asymptotic phase $\Phi_q({\bm \alpha})$ and amplitude $R_q({\bm \alpha})$ of the coherent state $\bm{\bm \alpha}$  since these functions are easily derived from the eigenoperators in numerical calculations.  
Additionally, we note that typical observables in quantum optics are often expressed in normal order, and their expectation values can be conveniently evaluated using the P-function \cite{carmichael2007statistical}.

\section{Numerical results}

To validate the proposed quantum asymptotic amplitude, we consider several representative 
quantum oscillatory systems that can introduce corresponding deterministic trajectories  in the classical limit.
First, we examine the quantum van der Pol oscillator with the quantum Kerr effect, which exhibits quantum limit-cycle oscillations in both the semiclassical and strong quantum regimes.
Next, we investigate the quantum van der Pol model with quantum squeezing, which exhibits quantum noise-induced oscillations near the saddle-node on invariant circle (SNIC) bifurcation in the classical limit \cite{strogatz1994nonlinear, guckenheimer1983nonlinear}, and a degenerate parametric oscillator with nonlinear damping, which exhibits quantum noise-induced oscillations near a pitchfork bifurcation in the classical limit \cite{strogatz1994nonlinear, guckenheimer1983nonlinear}.
We also analyze a harmonic oscillator with linear damping in Appendix \ref{sec:apdx_damp_ho}.

In our numerical calculations, we use the truncated matrix representation of the Liouville superoperator and compute the asymptotic phase and amplitude in Eqs.~(\ref{eq:phs_qt}) and~(\ref{eq:qiso}) from the eigensystem of this matrix \cite{kato2022definition}. 
When illustrating the quantum asymptotic phase, we select the sign of $\Omega_q$ such that  the asymptotic phase exhibits a counterclockwise increase. 

\subsection{Quantum van der Pol model with the quantum Kerr effect}
\label{sec:qvdp}

As a first example,  we consider the quantum van der Pol oscillator with the quantum Kerr effect.
The quantum master equation of the system is described by~\cite{lorch2016genuine, kato2022definition} 
\begin{align}
\label{eq:qvdp_me}
\dot{\rho} 
= {\mathcal L} \rho
= - i \left[ 
H
,\rho\right]
+ \gamma_{1} \mathcal{D}[a^{\dag}]\rho + \gamma_{2}\mathcal{D}[a^{2}]\rho,
\end{align}
where $a$ and $a^\dag$ denote the annihilation and creation operators, respectively. 
The Hamiltonian is defined as $H = \omega_0 a^{\dag}a + K a^{\dag 2} a^2$, where 
$\omega_{0}$ represents the frequency of the oscillator, 
$K$ is the Kerr parameter,
and $\gamma_{1}$ and $\gamma_{2}$ represent the decay rates for 
negative damping and nonlinear damping, respectively.
\subsubsection{Semiclassical regime}
\label{sec:qvdp_sc}

First, we consider the system in  the semiclassical regime where $\gamma_{2}$ and $K$ are sufficiently smaller than $\gamma_1$. In this regime, the quantum master equation in Eq.~(\ref{eq:qvdp_me}) is described by the quantum Fokker-Planck equation for $p({\bm \alpha})$ in Eq.~(\ref{eq:qpde}), and the system state is approximately described by a corresponding stochastic differential equation representing the deterministic trajectory subjected to the small quantum noise in the phase space ~\cite{kato2019semiclassical, kato2022definition}.

\begin{figure} [!t]
	\begin{center}
		\includegraphics[width=1\hsize,keepaspectratio]{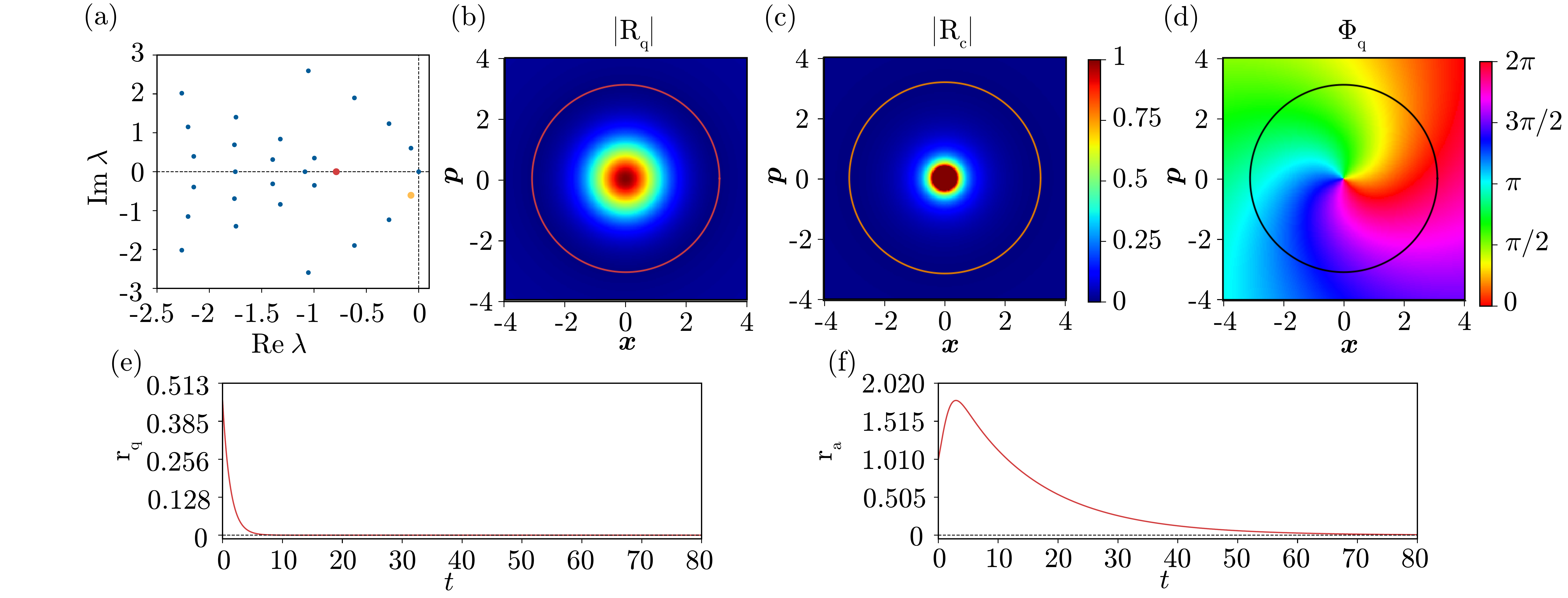}
	\end{center}
	\caption{
		Quantum asymptotic amplitude of a quantum van der Pol oscillator with the quantum Kerr effect in the semiclassical quantum regime, for $\gamma_1=1$ and $(\omega_0, \gamma_{2}, K)/\gamma_{1} = ({0.1}, 0.05, 0.025)$.
		(a) Eigenvalues of ${\mathcal L}^*$ in the vicinity of the imaginary axis. The red dot represents the largest non-zero real eigenvalue $\kappa_q$, and the yellow dot represents the principal eigenvalue $\overline{\Lambda_1}$.
		(b) Quantum asymptotic amplitude $\abs{R_q}$ with $\kappa_q = -0.787$.
		(c) Classical asymptotic amplitude $\abs{R_c}$ with $\kappa_c = -1$.
		(d) Quantum asymptotic phase $\Phi_q$ with $\Omega_q = -0.605$.
		(e-f) Evolution of $r_q$ and $r_a$:
		(e) $r_q $, (f) $r_a $.
		The red-thin line in (b) and the black-thin line in (d) represent the effective quantum periodic orbit $\chi_q$.
		The orange-thin line in (c) represents the deterministic limit cycle  $\chi_c$ in the classical limit.
		In (d),  the phase origin is set at $(x,p)=(2.5, 0)$.
	}
	\label{fig_1}
\end{figure}

Figure~\ref{fig_1}(a) shows the numerically obtained eigenvalues of ${\mathcal L}^*$ in the vicinity of the imaginary axis, where the real eigenvalue with the largest real part  $\kappa_q$ is marked with a red dot ($\kappa_q < 0$) and the eigenvalue $\overline{\Lambda_1}$ with the smallest absolute real part is marked with a yellow dot.
In this semiclassical regime, the eigenvalue $\overline{\Lambda_1} = \mu_q + i \Omega_q$  ($\mu_q < 0$) lies on the rightmost light-blue branch of the eigenvalues, which is approximately characterized by a parabola $\hat{\lambda}_n = i \Omega_q n + \mu_q n^2~ (n=0, \pm 1, \pm 2, \ldots)$ passing through $\overline{\Lambda_1}$. 
Therefore, we focus on the  slowest-decaying mode with the eigenvalue $\overline{\Lambda_1}$ because their imaginary parts are essentially integer multiples of $\Omega_q$. 

%$  $
In the classical limit, where quantum noise vanishes, the system is characterized by a complex variable $\alpha \in {\mathbb C}$, which evolves according to a deterministic  system derived from the drift term of the approximate quantum Fokker-Planck equation for $p({\bm \alpha})$
(see the details in Appendix B in Ref.~\cite{kato2022definition}),
\begin{align}
	\dot{\alpha} = \left( \frac{\gamma_1}{2} - i \omega_0 \right) \alpha 
	- (\gamma_{2}  + 2  K  i ) \overline{\alpha} \alpha^{2}.
	\label{eq:qvdp_ldv_alpha}
\end{align}

This model is known as the Stuart-Landau oscillator, the normal form of the supercritical Hopf bifurcation~\cite{kuramoto1984chemical}, and has a stable limit-cycle $\chi_c$. This limit-cycle attractor is explicitly calculated as $\alpha_c(\phi) = r_{c} e^{i \phi}$, where $\phi = \Omega_c t +const.$ with a natural frequency $\Omega_c = -\omega_0 - K \gamma_1/\gamma_2$ and $r_c = \sqrt{{\gamma_1} / {2 \gamma_{2} }}$
(accordingly, the quantum van der Pol model is also referred to as the quantum Stuart-Landau model in recent literature \cite{chia2020relaxation,mok2020synchronization}).
The basin of attraction of this limit cycle is the entire complex plane, excluding the origin.

\begin{figure} [htbp]
	\begin{center}
		\includegraphics[width=0.7\hsize,clip]{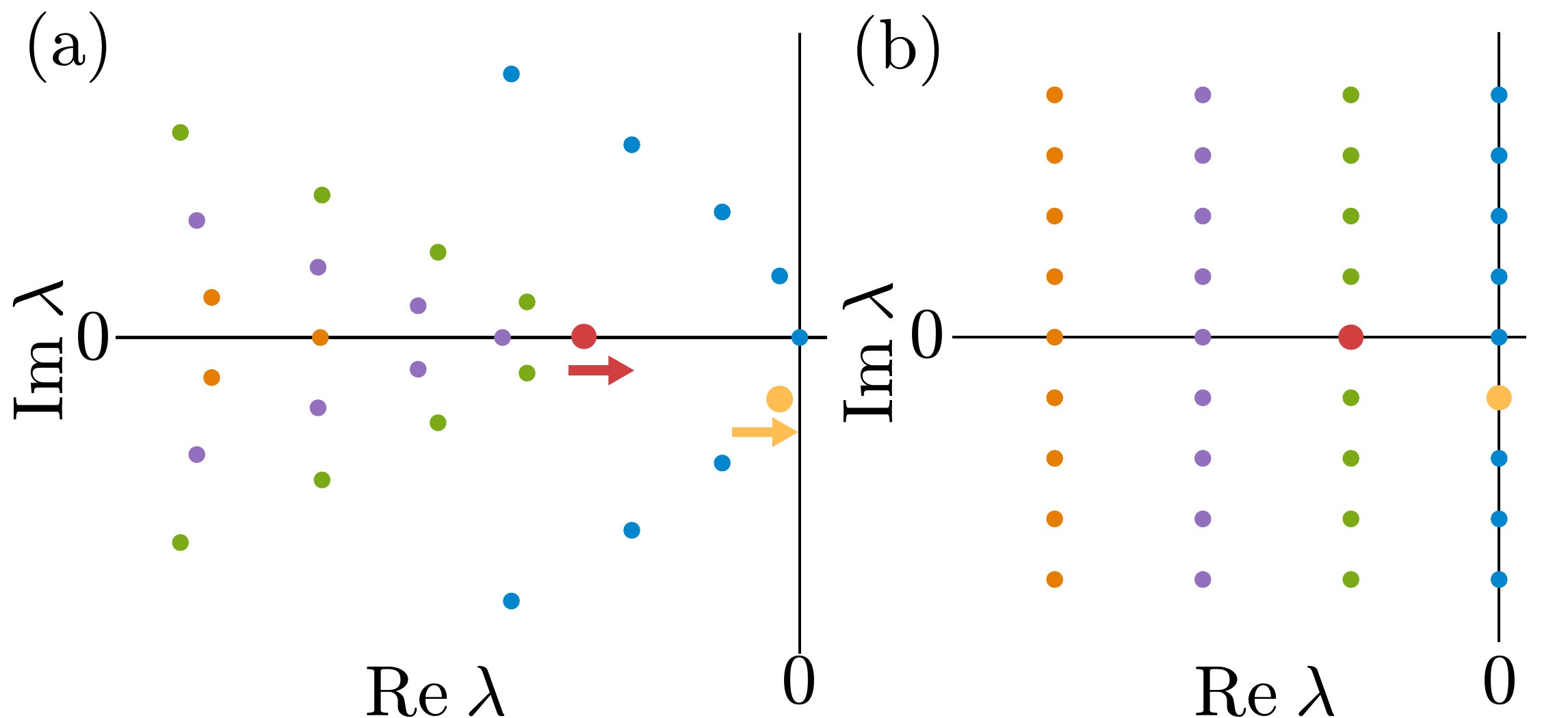}
		\caption{
			A schematic diagram illustrating eigenvalues of ${L}^{+}_{\bm \alpha}$ when approaching to the classical limit. (a) Semiclassical regime. (b) Classical limit. The red and yellow dots represent the eigenvalues related to the asymptotic amplitude and phase, respectively. 
		}
		\label{fig_2}
	\end{center}
\end{figure}

Figure~\ref{fig_2} presents a schematic diagram illustrating the eigenvalues of ${L}^{+}_{\bm \alpha}$, or ${\cal L}^{*}$, when the system approaches the deterministic system in the classical limit with a stable limit-cycle solution.
In the classical limit, the eigenvalues take the form $\lambda_c = m \kappa_c + i n \Omega_c~(m=0,1,2,\ldots$ and $n = 0, \pm 1, \pm2)$, with $\kappa_c =  -\gamma_1$ representing the largest non-zero real eigenvalue and $i \Omega_c$ representing the pure-imaginary eigenvalue with the smallest absolute imaginary part.
When the system approaches the deterministic system in the classical limit, the eigenvalues related to the asymptotic phase and amplitude, which appear in curved branches in the semiclassical regime in Fig.~\ref{fig_2}(a), converge to those in the corresponding straight branches of eigenvalues in the classical limit in Fig.~\ref{fig_2}(b).
The classical asymptotic amplitude $R_c$ of the system, associated with the eigenvalue $\kappa_c$ in the classical limit, is calculated as~\cite{kato2021asymptotic}
\begin{align}
\label{eq:qvdp_iso}
R_c({\bm \alpha}) = c_0 \left( \gamma_2 - \frac{\gamma_1}{2 (x^2 + p^2)} \right),
\end{align}
where $(x, p) = ( \mbox{Re}~\alpha, \mbox{Im}~\alpha )$ and 
$c_0$ is an arbitrary real constant because the scale of the asymptotic amplitude can be freely chosen.
The amplitude satisfies $\dot{R}_c({\bm \alpha})  = \kappa_c R_c({\bm \alpha})$ when ${\alpha}$ evolves within the basin of attraction of the limit cycle under Eq.~(\ref{eq:qvdp_ldv_alpha}).

Figures~\ref{fig_1}(b) and~\ref{fig_1}(c) illustrate the quantum asymptotic amplitude $\abs{R_q( \bm \alpha)}$ and the corresponding asymptotic amplitude $\abs{R_c({\bm \alpha})}$, respectively.  Both  amplitude functions are scaled by a constant factor such that the maximum value of $\abs{R_q}$ becomes $1$, and the same color map is used in both figures. In Fig.~\ref{fig_1}(c),  values of $\abs{R_c}$ greater than $1$ near the origin are illustrated in the same color. 
 In Fig.~\ref{fig_1}(b),  the red-thin line illustrates the effective quantum periodic orbit 
 $\chi_q$, and in Fig.~\ref{fig_1}(c), the orange-thin line illustrates the deterministic limit cycle $\chi_c~(= \{ \bm{\alpha} | R_c(\bm \alpha) = 0 \})$ 
 in the classical limit. 
 The quantum asymptotic amplitude $R_q(\rho)~(= \langle  p({\bm \alpha}), { R_q({\bm \alpha}) } \rangle_{\bm \alpha})$ of the quantum state $\rho$ decays exponentially to zero, 
 indicating that the system evolves toward the vicinity of the effective quantum periodic orbit 
 $\chi_q$. In the classical limit, the quantum periodic orbit $\chi_q$ coincides with the classical limit cycle $\chi_c$. Therefore, in this regime with sufficiently small quantum noise,  the quantum periodic orbit $\chi_q$ closely matches the classical limit-cycle $\chi_c$.  

In Fig.~\ref{fig_1}(d), we also plot the quantum asymptotic phase $\Phi_q({\bm \alpha})$ together with the effective quantum periodic orbit $\chi_q$.  The phase increases in the  counterclockwise direction along the quantum periodic orbit. 

To confirm that the quantum asymptotic amplitude produces appropriate amplitude values, we numerically evaluate the free relaxation of $\rho$ starting from an initial coherent state  $\rho_0 = \ket{\alpha_0}\bra{\alpha_0}$ with $\alpha_0=1$ at $t=0$. We evaluate $r_q = \abs{R_q(\rho)} =  \abs{\langle \rho, V_2 \rangle_{tr}}$ of the eigenoperator $V_2$ and compare it with $r_a = \abs{\langle a \rangle_q } = \abs{\langle \rho, a \rangle_{tr}}$ of the annihilation operator $a$, which represents the polar radius of $\langle a \rangle$ on the complex plane.  Due to dissipation, the system evolves from the initial pure coherent state into a mixed state and eventually reaches a stationary mixed state. Figures~\ref{fig_1}(e) and~\ref{fig_1}(f) depict the time evolution of $r_q$ and $r_a$, respectively. The asymptotic amplitude $r_q$ decays exponentially with a constant rate $\kappa_q$ in Fig.~\ref{fig_1}(e). In contrast, the radius $r_a$ does not exhibit an exponential decay over time; particularly, it increases during the transient evolution before $t=5$ in Fig.~\ref{fig_1}(f). 
These results confirm that, in the semiclassical regime, the quantum asymptotic amplitude  provides an appropriate and consistent amplitude values.

\subsubsection{Strong quantum regime}
\label{sec:qvdp_qm}

Next, we consider the strong quantum regime, where $\gamma_{2}$ and $K$ are relatively large compared to $\gamma_1$. In this regime, the system dynamics 
is dominated by a small number of energy states and the semiclassical approximation becomes inapplicable.

\begin{figure} [!t]
	\begin{center}
		\includegraphics[width=1\hsize,clip]{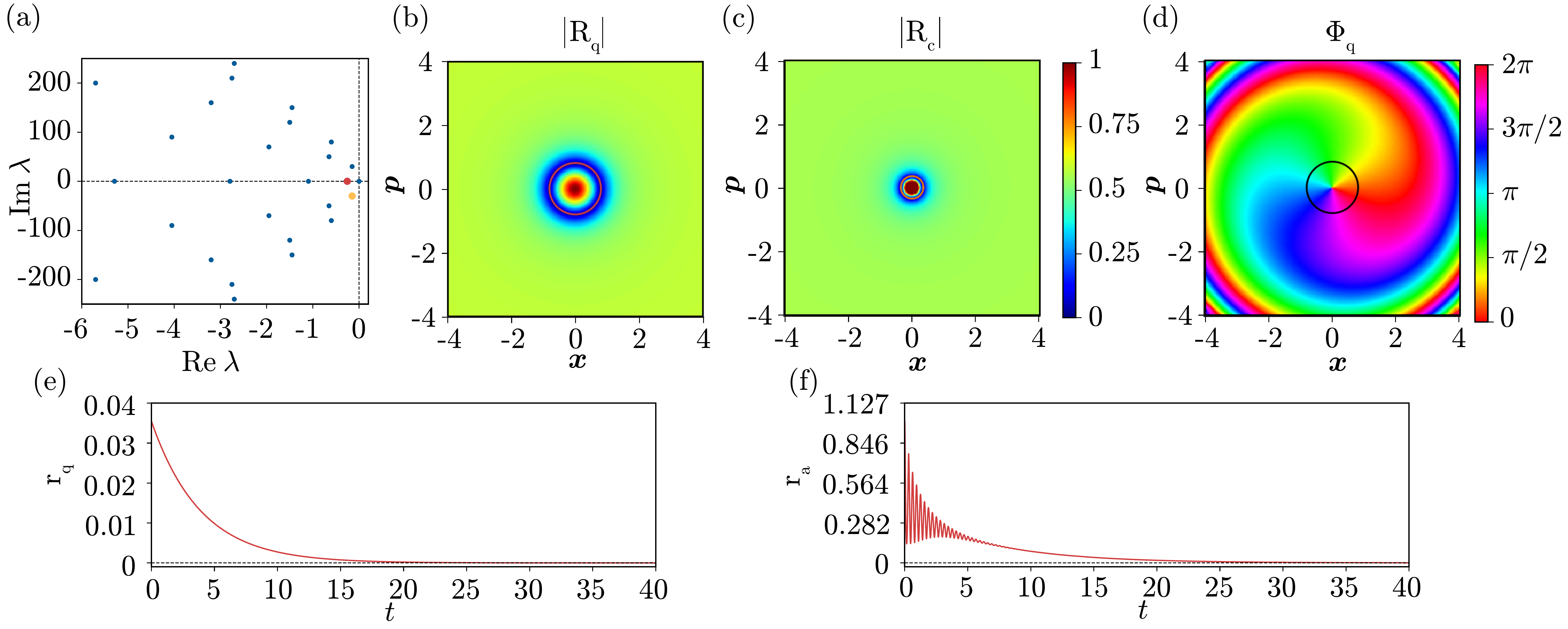}
		\caption{
			Quantum asymptotic amplitude of a quantum van der Pol oscillator with the quantum Kerr effect in the strong quantum regime, for 
			$\gamma_1=0.1$ and $(\omega_0, \gamma_{2}, K)/\gamma_{1} = (300, 4, 100)$.
			(a) Eigenvalues of ${\mathcal L}^*$ in the vicinity of the imaginary axis. The red dot represents the largest non-zero real eigenvalue $\kappa_q$, and the yellow dot represents the principal eigenvalue $\overline{\Lambda_1}$.
			(b) Quantum asymptotic amplitude $\abs{R_q}$ with $\kappa_q = -0.256$.
			(c) Classical asymptotic amplitude $\abs{R_c}$ with $\kappa_c = -0.1$.
			(d) Quantum asymptotic phase $\Phi_q$ with $\Omega_q = -30$.
			(e-f) Evolution of $r_q$ and $r_a$:
			(e) $r_q $, (f) $r_a $.
			The red-thin line in (b) and the black-thin line in (d) represent the effective quantum periodic orbit $\chi_q$.
			The orange-thin line in (c) represents the deterministic limit cycle  $\chi_c$ in the classical limit.
			In (d),  the phase origin is set at $(x,p)=(2.5, 0)$.
		}
		\label{fig_3}
	\end{center}
\end{figure}

Figure~\ref{fig_3}(a) shows the numerically obtained eigenvalues of ${\mathcal L}^*$ in the vicinity of the imaginary axis, with $\kappa_q$ marked with a red dot and  $\overline{\Lambda_1}$ marked with a yellow dot.

Figures~\ref{fig_3}(b) and~\ref{fig_3}(c) depict the quantum asymptotic amplitude $\abs{R_q( \bm \alpha)}$ and the corresponding asymptotic amplitude $\abs{R_c({\bm \alpha})}$, respectively, where both  amplitude functions are scaled by a constant factor such that the maximum value of $\abs{R_q}$ becomes $1$, the same color map is applied to both figures, and  values of $\abs{R_c}$ exceeding $1$ near the origin in Fig.~\ref{fig_3}(c) are illustrated in the same color. 
The effective quantum periodic orbit $\chi_q $ and the deterministic limit cycle $\chi_c$ in the classical limit are plotted by the red-thin line in Fig.~\ref{fig_3}(b) and the orange-thin line in Fig.~\ref{fig_3}(c), respectively.  In the strong quantum regime,  the quantum periodic orbit $\chi_q$ clearly deviates from the classical limit cycle $\chi_c$.

In Fig.~\ref{fig_3}(d), we also plot the quantum asymptotic phase $\Phi_q({\bm \alpha})$ together with the effective quantum periodic orbit $\chi_q$, where the phase increases  in the counterclockwise direction along the quantum periodic orbit. 

We numerically evaluate the free relaxation of $\rho$, starting from an initial coherent state $\rho_0 =  \ket{\alpha_0}\bra{\alpha_0}$ with $\alpha_0=1$ at $t=0$, and plot the time evolution of the asymptotic amplitude $r_q$ and radius $r_a$. 
Figures~\ref{fig_3}(e) and (f) depict the time evolution of $r_q$ and $r_a$, respectively. As expected, the asymptotic amplitude $r_q$ exhibits an exponential decay with a constant rate $\kappa_q$. In contrast, the radius $r_a$ does not decay exponentially but instead oscillates rapidly over time. 
These results indicate that, despite strong quantum effects significantly altering the system dynamics from the classical limit, the quantum asymptotic amplitude $R_q$  reliably produces isostable amplitude values even in the strong quantum regime.

\subsection{Quantum van der Pol model with the quantum squeezing}
\label{sec:qvdp_sq}
Next, we consider a system exhibiting quantum noise-induced oscillations, i.e., oscillatory responses excited by quantum noise \cite{kato2021quantum}.
Specifically, we consider a system in the classical limit lies near an SNIC bifurcation \cite{strogatz1994nonlinear, guckenheimer1983nonlinear}.
As a minimum model exhibiting such quantum noise-induced oscillations, we consider a quantum van der Pol model subjected to the quantum squeezing~\cite{sonar2018squeezing}.
In the classical limit,  the corresponding deterministic system behaves as an excitable bistable system slightly below the bifurcation point of spontaneous limit-cycle oscillations
\cite{kato2021quantum}.
Let $\omega_{sq}$ denote the frequency of the pump beam of the quantum squeezing.  
In the rotating coordinate frame with frequency $\omega_{sq}/2$,  the system evolves according to  the following quantum master equation
\cite{sonar2018squeezing, kato2019semiclassical, kato2021quantum}
\begin{align}
	\label{eq:qvdp_sq_me}
	\dot{\rho} = 
	-i \left[  - \Delta a^{\dag}a 
	+ i \eta ( a^2 e^{-i \theta} - a^{\dag 2} e^{ i \theta}  )
	, \rho \right]
	+ \gamma_{1} \mathcal{D}[a^{\dag}]\rho + \gamma_{2}\mathcal{D}[a^{2}]\rho,
\end{align}
where, $\Delta = \omega_{sq}/2 - \omega_{0}$ denotes the detuning of
the half frequency of the pump beam of squeezing from the system's oscillatory frequency, and $\eta e^{ i \theta}$ ($\eta \geq 0$, $0 \leq \theta < 2\pi$) denotes the squeezing parameter.

In the classical limit, where the quantum noise vanishes, the system is described by a complex variable $\alpha \in {\mathbb C}$, which evolves according to  the deterministic system  (see the derivation in Appendix \ref{sec:apdx_uni})
\begin{align}
	\dot{\alpha} = \left( \frac{\gamma_1}{2} + i \Delta \right) \alpha - \gamma_{2} \overline{\alpha} \alpha^{2} - 2 \eta e^{i \theta} \overline{\alpha}
	\label{eq:qvdp_sq_ldv_alpha}.
\end{align}

By using the modulus $R$ and argument $\phi$ of the complex variable $\alpha = R e^{i \phi}$, 
 the differential equations for these variables are obtained by
\begin{align}
	\dot{R} &=  \frac{\gamma_1}{2} R-\gamma_{2} R^{3} - 2 \eta R \cos (2 \phi-\theta), 
	\label{eq:qvdp_sq_ldv_rad}
	\\
	\dot{\phi} &= \Delta + 2 \eta \sin (2 \phi-\theta). 
	\label{eq:qvdp_sq_ldv_phi}
\end{align}
When the squeezing effect is present, i.e., $\eta \neq 0$, the system has two stable fixed points for $\Delta \leq 2\eta$, as seen in the equation for the argument $\phi$ in Eq.~(\ref{eq:qvdp_sq_ldv_phi}). 
These fixed points annihilate with their corresponding unstable fixed points via an SNIC bifurcation at $\Delta = 2 \eta$;  a stable limit-cycle emerges and the argument $\phi$ continuously increases over time when $\Delta > 2\eta$, whereas $\phi$ has two stable fixed points and converges to one of them depending on the initial state when $\Delta < 2\eta$.
When the system in the classical limit is slightly below the onset of the SNIC bifurcation, namely, when $\Delta$ is slightly less than $2\eta$, oscillatory response can be induced by the quantum noise. In what follows, we consider such a parameter configuration (see Appendix \ref{sec:apdx_uni} for details).

\begin{figure} [!t]
	\begin{center}
		\includegraphics[width=1\hsize,keepaspectratio]{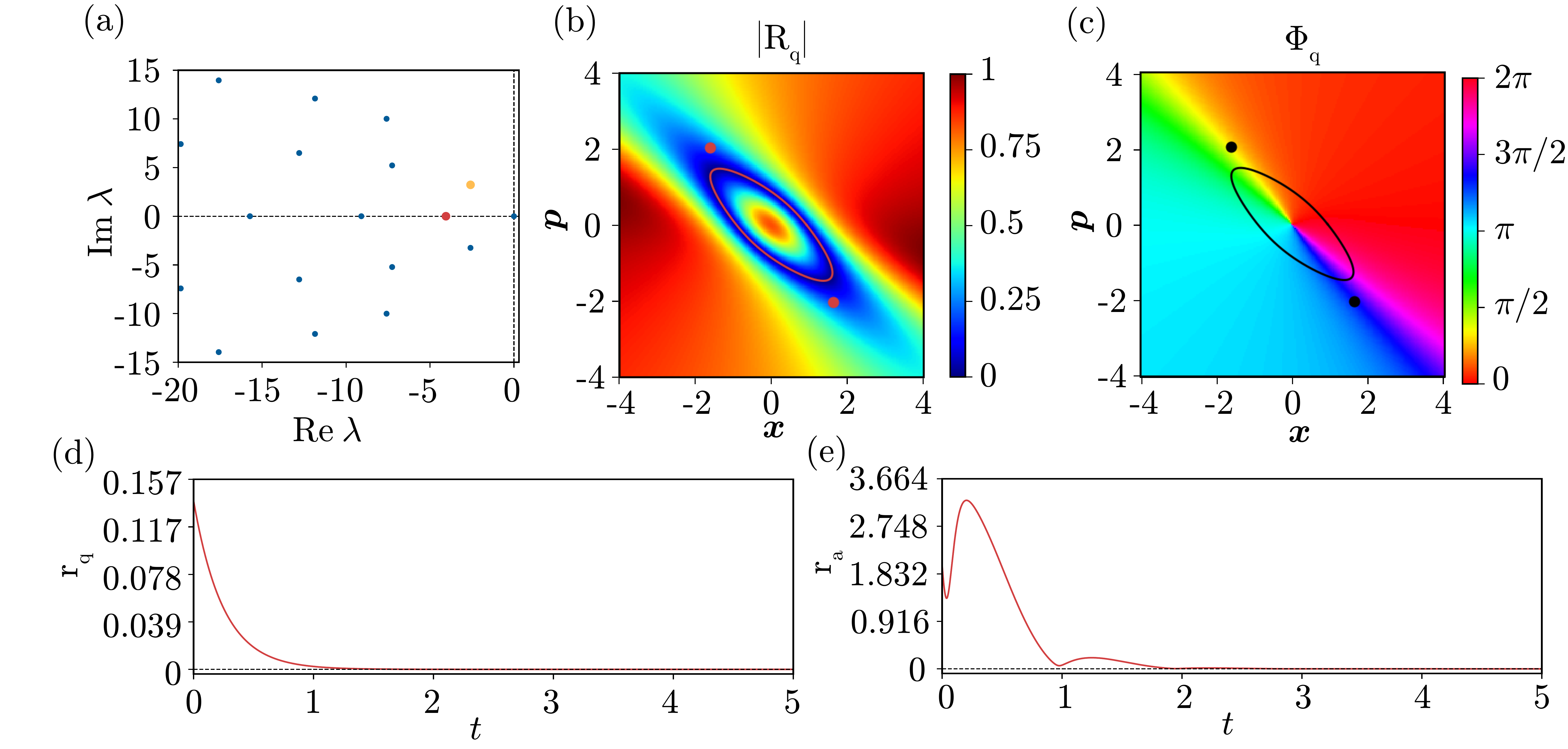}
	\end{center}
	\caption{
		Quantum asymptotic amplitude of a quatnumin van der Pol model with the quantum squeezing effect, for $\gamma_1=1$ and $(\Delta, \gamma_{2}, \eta)/\gamma_{1} = (13.65, 0.53, 7)$.
		(a) Eigenvalues of ${\mathcal L}^*$ in the vicinity of the imaginary axis. The red dot represents the largest non-zero real eigenvalue $\kappa_q$, and the yellow dot represents the principal eigenvalue $\overline{\Lambda_1}$.
		(b) Quantum asymptotic amplitude $\abs{R_q}$ with $\kappa_q = -4.052$.
		(c) Quantum asymptotic phase $\Phi_q$ with $\Omega_q = 3.248$.
		(d,e) Evolution of $r_q$ and $r_a$:
		(d) $r_q $, (e) $r_a $.
		The red-thin line in (b) and the black-thin line in (c) represent the effective quantum periodic orbit $\chi_q$, and the two red dots in (b) and the two black dots in (c) represent the two stable fixed points in the classical limit.
		In (c),  the phase origin is set at $(x,p)=(2.5, 0)$.
	}
	\label{fig_4}
\end{figure}

Figure~\ref{fig_4}(a) shows the numerically obtained eigenvalues of ${\mathcal L}^*$ in the vicinity of the imaginary axis,  with $\kappa_q$ marked with a red dot and  $\overline{\Lambda_1}$ marked with a yellow dot.
Figures~\ref{fig_4}(b) and~\ref{fig_4}(c) depict the quantum asymptotic amplitude $\abs{R_q( \bm \alpha)}$ and phase $\Phi_q( \bm \alpha)$, respectively, where the amplitude function is scaled by a constant factor such that the maximum value of $\abs{R_q}$ becomes $1$. 
Although the system in the classical limit is set below the SNIC bifurcation and only possesses two stable fixed points, which are shown by the two red dots in Fig.~\ref{fig_4}(b) and two black dots in Fig.~\ref{fig_4}(c), the oscillatory response excited by the quantum noise gives rise to an effective quantum periodic orbit $\chi_q$, which is illustrated by the red-thin line in Fig.~\ref{fig_4}(b) and black-thin line in Fig.~\ref{fig_4}(c).  
%%%
This effective quantum periodic orbit is illustrated using the quantum asymptotic amplitude and  cannot be introduced from the deterministic dynamics of the system in the classical limit. 

The quantum asymptotic phase $\Phi_q({\bm \alpha})$ can also be introduced using Eq.~(\ref{eq:phs_qt}),  which increases in the counterclockwise direction along the effective quantum periodic orbit $\chi_q$, as depicted in Fig.~\ref{fig_4}(c). 
The quantum asymptotic phase captures the underlying fast-slow dynamics corresponding to the round trip between two stable fixed points, with slow dynamics (rapid phase change) near the two stable fixed points and fast dynamics (moderate phase change) between them, as depicted in Fig.~\ref{fig_4}(c).  It should be noted that this is the first illustration of an asymptotic phase for a quantum noise-induced oscillation.

We consider the free relaxation of $\rho$ from an initial coherent state $\rho_0 =  \ket{\alpha_0}\bra{\alpha_0}$ with $\alpha_0=2$ at $t=0$ and plot the time evolution of the asymptotic amplitude $r_q$ and the radius $r_a$. Figures~\ref{fig_4}(d) and (e) depict the time evolution of $r_q$ and $r_a$, respectively. As expected, the asymptotic amplitude $r_q$ decays exponentially with a constant rate $\kappa_q$. In contrast, the radius $r_a$ does not exhibit an exponential decay over time; instead, it increases during the transient evolution before $t=2$.

\subsection{A degenerate parametric oscillator with nonlinear damping}
\label{sec:dpo}
Next, we consider another system exhibiting quantum noise-induced oscillations; particularly, we consider a system in the classical limit to be near a pitchfork bifurcation \cite{strogatz1994nonlinear, guckenheimer1983nonlinear}.
As a minimum model exhibiting such quantum noise-induced oscillations, we consider a degenerate parametric oscillator with nonlinear damping \cite{tezak2017low, kato2022turing}, whose deterministic system in the classical limit can exhibit a pitchfork bifurcation.

The pump beam for the squeezing is also added in this model. In the rotating coordinate frame with a frequency of $\omega_{sq}/2$, the system dynamics are governed by 
the following quantum master equation~\cite{tezak2017low, kato2022turing}
\begin{align}
	\dot{\rho} = 
	-i \left[ -\Delta a^{\dag}a 
	+ i \eta ( a^2 e^{-i \theta} - a^{\dag 2} e^{ i \theta}  )
	, \rho \right]
	+ \gamma_{2}\mathcal{D}[a^{2}]\rho + \gamma_{3} \mathcal{D}[a]\rho,
	\label{eq:dpo_me}
\end{align}
where $\gamma_3$ represents the decay rate for linear damping. 

In the classical limit, where the quantum noise vanishes,  the system is characterized by a single complex variable $\alpha \in {\mathbb C}$, which evolves according to  a deterministic system  (see the derivation in Appendix \ref{sec:apdx_uni})
\begin{align}
	\dot{\alpha} = \left( \frac{-\gamma_3}{2} + i \Delta \right) \alpha - \gamma_{2} \overline{\alpha} \alpha^{2} - 2 \eta e^{i \theta} \overline{\alpha}
	\label{eq:qvdp_sq_ldv_alpha}.
\end{align}

Using the modulus $R$ and argument $\phi$ of the complex variable $\alpha = R e^{i \phi}$, the differential equations for these variables are given by
\begin{align}
\label{eq:dpo_ldv_rad}
\dot{R} &=  \frac{- \gamma_3}{2} R-\gamma_{2} R^{3} - 2 \eta R \cos (2 \phi-\theta), 
\\
\label{eq:dpo_ldv_phi}
\dot{\phi} &= \Delta + 2 \eta \sin (2 \phi-\theta). 
\end{align}
When the squeezing effect is present, i.e., $\eta \neq 0$, the equation for the argument in Eq.~(\ref{eq:dpo_ldv_phi}), possesses two stable fixed points for $\Delta \leq 2\eta$.
In this case, a non-zero positive solution of the equation for the modulus in Eq.~(\ref{eq:dpo_ldv_rad}) arises via a pitchfork bifurcation at $\eta = \frac{1}{2} \sqrt{ \frac{\gamma^2_3}{4} + \Delta^2} $;  the system possesses two stable fixed points with a common modulus when $\eta > \frac{1}{2} \sqrt{ \frac{\gamma^2_3}{4} + \Delta^2}$, while it possesses a single stable fixed point at the origin when  $\eta \leq \frac{1}{2} \sqrt{ \frac{\gamma^2_3}{4} + \Delta^2}$.

We assume that the equation for the modulus in Eq.~(\ref{eq:dpo_ldv_rad}) is slightly below the bifurcation, namely 
$\eta < \frac{1}{2} \sqrt{ \frac{\gamma^2_3}{4} + \Delta^2}$, and the equation for the argument in Eq.~(\ref{eq:dpo_ldv_phi}) is slightly below the bifurcation, namely  $\Delta < 2\eta$.  In this case,  due to the effect of quantum noise, 
the system can exceed the pitchfork bifurcation point, leading to the emergence of two stable fixed points. Simultaneously, it exceeds the bifurcation point $\Delta = 2\eta$ of the argument and exhibits an oscillatory response characterized by the round trip around two emerging stable fixed points, resulting in a quantum noise-induced oscillation starting from the origin.  In what follows, we consider such a parameter configuration (see the details in Appendix \ref{sec:apdx_uni}).

\begin{figure} [!t]
	\begin{center}
		\includegraphics[width=1\hsize,keepaspectratio]{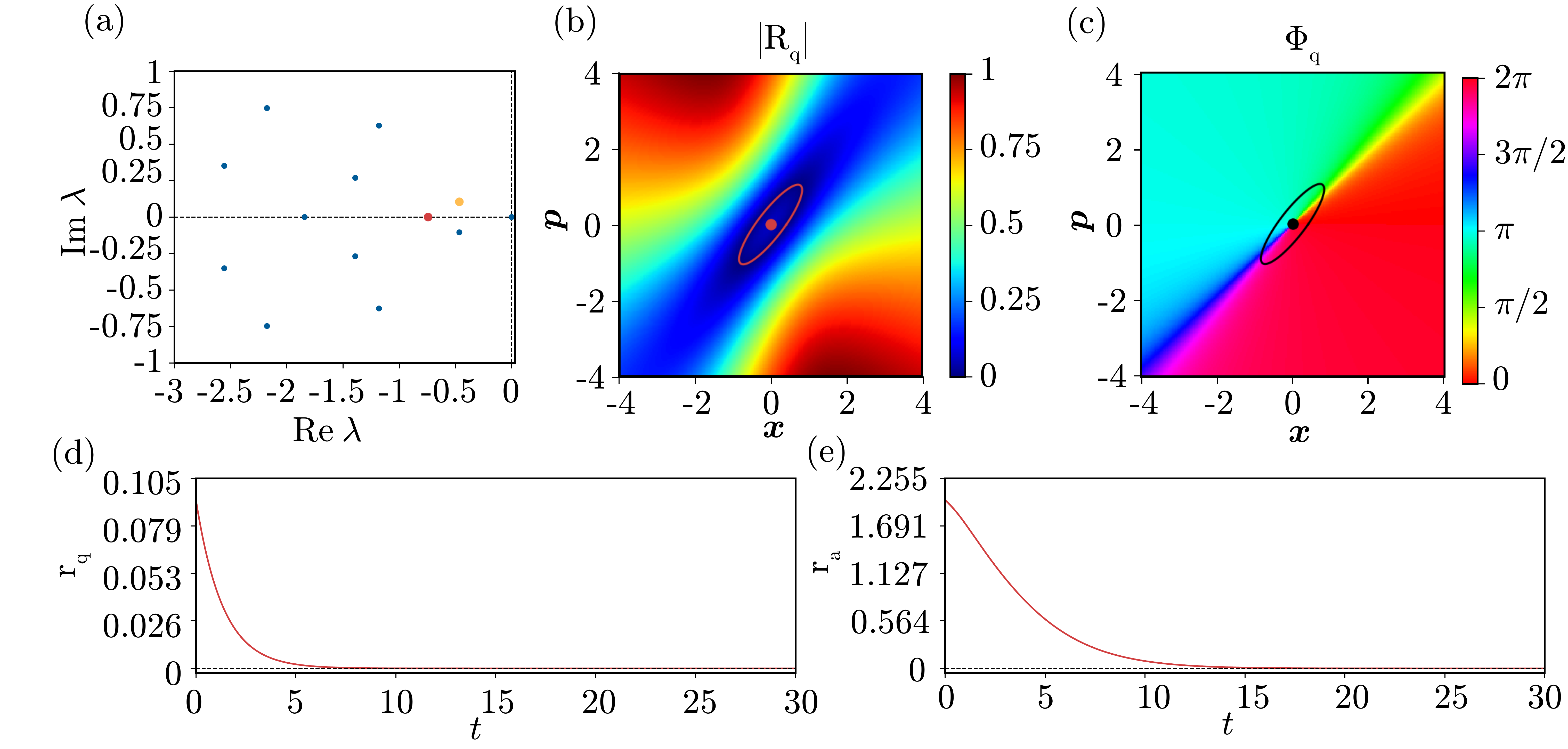}
	\end{center}
	\caption{
		Quantum asymptotic amplitude of a degenerate parametric oscillator with nonlinear damping, for $\gamma_2=0.1$ and $(\Delta, \gamma_{3},  \eta e^{-i \theta})/\gamma_{2} = (6, 7.5, -3.25)$.
		(a) Eigenvalues of ${\mathcal L}^*$ in the vicinity of the imaginary axis. The red dot represents the largest non-zero real eigenvalue $\kappa_q$, and the yellow dot represents the principal eigenvalue $\overline{\Lambda_1}$.
		(b) Quantum asymptotic amplitude $\abs{R_q}$ with $\kappa_q = -0.744$.
		(c) Quantum asymptotic phase $\Phi_q$ with $\Omega_q = 0.105$.
		(d,e) Evolution of $r_q$ and $r_a$:
		(d) $r_q $, (e) $r_a $.
		The red-thin line in (b) and the black-thin line in (c) represent the effective quantum periodic orbit $\chi_q$, and the red dot in (b) and the black dot in (c) represent the stable fixed point at the origin in the classical limit.
		In (c),  the phase origin is set at $(x,p)=(2.5, 0)$.
}
\label{fig_5}
\end{figure}

Figure~\ref{fig_5}(a) shows the numerically obtained eigenvalues of ${\mathcal L}^*$ in the vicinity of the imaginary axis, with $\kappa_q$ marked with a red dot and  $\overline{\Lambda_1}$ marked with a yellow dot.

Figures~\ref{fig_5}(b) and~\ref{fig_5}(c) depict the quantum asymptotic amplitude $\abs{R_q( \bm \alpha)}$ and phase $\Phi_q( \bm \alpha)$, respectively, where the amplitude function is scaled by a constant factor such that the maximum value of $\abs{R_q}$ becomes $1$. Although the system in the classical limit is set below a pitchfork bifurcation and has only a stable fixed point at the origin, which is shown by the red dot in Fig.~\ref{fig_5}(b) and black dot in Fig.~\ref{fig_5}(c), the oscillatory response excited by the quantum noise leads to the emergence of an effective quantum periodic orbit $\chi_q$, which is illustrated by the red-thin line in Fig.~\ref{fig_5}(b) and black-thin line in Fig.~\ref{fig_5}(c).  
%%%
This effective quantum periodic orbit is illustrated using the proposed quantum asymptotic amplitude and cannot be introduced from the deterministic dynamics of the system in the classical limit. 

The quantum asymptotic phase $\Phi_q({\bm \alpha})$ can also be introduced using Eq.~(\ref{eq:phs_qt}),  which increases in the counterclockwise direction along the effective quantum periodic orbit $\chi_q$, as depicted in Fig.~\ref{fig_5}(c). 
The quantum asymptotic phase reflects fast-slow dynamics, similar to the behavior observed in Fig.~\ref{fig_4}(c), which corresponds to the nose-induced oscillations starting from the origin.

We consider the free relaxation of $\rho$ from an initial coherent state $\rho_0 =  \ket{\alpha_0}\bra{\alpha_0}$ with $\alpha_0=2$ at $t=0$ and plot the time evolution of the asymptotic amplitude $r_q$ and the radius $r_a$. Figures~\ref{fig_5}(d) and (e) illustrate the time evolution of $r_q$ and $r_a$, respectively. As expected, the asymptotic amplitude $r_q$ decays exponentially with a constant rate $\kappa_q$. In contrast, the radius $r_a$ does not exhibit an exponential decay over time.
Note that similar noise-induced oscillations can also arise in classical stochastic systems when its noiseless deterministic dynamics lies slightly below a pitchfork bifurcation. In analogy with previous studies on quasicycles occurring below a supercritical Hopf bifurcation in the corresponding deterministic systems~\cite{brooks2015quasicycles}, these oscillations can be interpreted as quasicycles emerging below a pitchfork bifurcation.

\subsection{Discussion}

In this study, we focused on asymptotic amplitude functions for quantum oscillatory systems exhibiting quantum limit-cycle oscillations and quantum noise-induced oscillations. In classical deterministic systems, asymptotic amplitudes, also referred to as isostables, can be defined even for linear systems with stable fixed points~\cite{mauroy2013isostables, mauroy2020koopman}. Similarly, the quantum asymptotic amplitude can be introduced for quantum oscillatory systems, even when their classical counterparts do not exhibit, or are not near, bifurcations associated with nonlinear oscillations. As demonstrated in Appendix~\ref{sec:apdx_damp_ho}, we explicitly compute the asymptotic amplitude function for a quantum harmonic oscillator with linear damping, a linear system that does not exhibit nonlinear oscillatory behavior.

 For deterministic classical nonlinear systems possessing a stable limit-cycle solution,  the zero-level set of the asymptotic amplitude function coincides with the stable limit cycle $\chi_c$. Hence, the effective quantum periodic orbit $\chi_q$, which is defined as the zero-level set of the quantum asymptotic amplitude, can be regarded as a natural generalization of the limit-cycle attractor for quantum nonlinear oscillatory systems. As shown in the examples in Sec. \ref{sec:qvdp_qm}, this effective quantum periodic orbit can also be introduced even in the strong quantum regime and may be used for formulating a phase reduction theory for quantum synchronization beyond the semiclassical regime. Moreover, we can also introduce the effective quantum periodic orbit in examples of quantum noise-induced oscillations in Sec. \ref{sec:qvdp_sq} and \ref{sec:dpo}. This effective quantum periodic orbit cannot be derived without introducing the quantum asymptotic amplitude and offers intuitive  insights into the dynamics of quantum oscillatory systems in the phase space of quasiprobability distribution.

The asymptotic amplitude characterizes the magnitude of the slowest decaying mode and captures the slowest asymptotic approach of the system to its steady state. Accordingly, it differs from the conventional notion of amplitude, i.e., the radius,  associated with oscillators. Moreover, the asymptotic amplitude is conceptually distinct from the amplitude equations used in pattern formation theory~\cite{cross1993pattern}, which describe the dynamics of unstable modes near bifurcation points through differential equations for complex variables.

In this study, our focus is solely on the eigenfunction associated with the largest non-zero real eigenvalue to introduce the quantum asymptotic amplitude. Recently, regarding the slowest-decaying eigenfunction of the backward Fokker-Planck (Kolmogorov) operator, not only the argument used for defining the stochastic asymptotic phase, but also the modulus of the eigenfucntion is used for formulating a universal description of stochastic oscillators \cite{perez2023universal}. 
Moreover,  eigenfunctions associated with dominant eigenvalues from different eigenvalue branches can be used to introduce multiple quantum asymptotic phases for analyzing quantum synchronization in the strong quantum regimes \cite{kato2023quantum}. Thus, besides the eigenfunctions associated with the largest non-zero real eigenvalue, the modulus of the other dominant eigenfunctions, which also exhibit exponential decay,  may also provide valuable information for analyzing quantum oscillatory systems in the strong quantum regime.

\section{Conclusions}

We introduced a definition of the asymptotic amplitude for quantum oscillatory systems by extending the asymptotic amplitude for classical stochastic oscillatory systems from the Koopman operator viewpoint~\cite{kato2021asymptotic}. The quantum asymptotic amplitude produces appropriate isostable amplitude values and introduces effective quantum periodic orbits for quantum limit-cycle oscillations and quantum noise-induced oscillations. 
Despite increasing interest in quantum synchronization, the theoretical tools for analyzing quantum rhythmic oscillations and synchronization remain less developed than those for classical systems. 
When combined with the quantum asymptotic phase introduced in our previous work~\cite{kato2022definition}, the proposed quantum asymptotic amplitude, applicable even in strong quantum regime,  may offer novel tools for qualitatively analyzing quantum synchronization.

In this study, we focused on quantum nonlinear oscillators in quantum optical systems due to their clear correspondence with classical nonlinear oscillatory systems.
Recently, quantum limit-cycle oscillations in quantum spin systems have been investigated through their corresponding deterministic dynamics in the classical limit~\cite{dutta2025quantum}. In particular, we have recently proposed the quantum spin van der Pol oscillator, which exhibits a classical limit-cycle trajectory corresponding to the normal form of a supercritical Hopf bifurcation~\cite{kato2024quantum}.
Therefore,  as a future study, our definitions of quantum asymptotic phase and amplitude functions also need to be investigated for quantum oscillatory systems in the quantum spin systems and may be used for analyzing quantum spin synchronization \cite{roulet2018synchronizing, roulet2018quantum}. 

Recently, a phase reduction theory for quantum nonlinear oscillators, applicable even in the strong quantum regime, has been developed based on the stochastic Schr\"odinger equation. This formulation relies on the limit-cycle trajectory reconstructed via quantum continuous measurement~\cite{setoyama2024lie}.
By utilizing the proposed quantum asymptotic phase, amplitude, and effective quantum periodic orbit, it may be possible to formulate a phase-amplitude reduction theory for quantum nonlinear oscillators directly from the quantum master equation. Such a framework would reduce the full quantum dynamics into phase and amplitude equations, thereby simplifying the analysis of quantum oscillatory behavior and enabling a systematic analysis of quantum synchronization in the strong quantum regime.

Beyond the analysis of quantum nonlinear oscillatory systems, the Koopman operator perspective may open a new avenue for studying open quantum systems. Notably, Koopman operator theory for classical Hamiltonian systems was originally inspired by the operator formalism of quantum Hamiltonian systems \cite{koopman1931hamiltonian}. With recent progress in applying Koopman operator theory to classical dissipative systems~\cite{mauroy2020koopman}, extending this framework to quantum dissipative, or open quantum, systems presents a promising direction for the detailed analysis, control, and optimization of complex quantum nonlinear dynamics.

\begin{acknowledgements}
	The authors gratefully acknowledge stimulating discussions with Hiroya Nakao.
	Numerical simulations have been performed by using QuTiP numerical
	toolbox~\cite{johansson2012qutip,johansson2013qutip}. 
	We acknowledge JSPS KAKENHI JP22K14274, JST PRESTO JP-MJPR24K3 and JST CREST JP-MJCR1913 for financial support. 
\end{acknowledgements}

\subsection*{Conflict of Interest}
The authors have no conflicts to disclose.

\subsection*{Data availability}
The data that supports the findings of this study are available within the article.

\appendix

\section{Models exihibiting noise-induced oscillaitons in the semiclassical regime}
\label{sec:apdx_uni}

In this section, we briefly explain the classical limit of the models exhibiting quantum noise-induced oscillations discussed in Sec. \ref{sec:qvdp_sq}, and \ref{sec:dpo}. To present these models in a unified manner, 
%a
we consider the following quantum master equation,
\begin{align}
	\label{eq:uni_me}
	\dot{\rho} = 
	-i \left[  -\Delta a^{\dag}a 
	+ i \eta ( a^2 e^{-i \theta} - a^{\dag 2} e^{ i \theta}  )
	, \rho \right]
	+ \gamma_{1} \mathcal{D}[a^{\dag}]\rho + \gamma_{2}\mathcal{D}[a^{2}]\rho + \gamma_{3} \mathcal{D}[a]\rho,
\end{align}

In the semiclassical regime, the linear operator $L_{\bm \alpha}$ in Eq.~(\ref{eq:qpde}) can be described by a Fokker-Planck operator, 
\begin{align}
	L_{\bm \alpha} = \Big[ - \sum_{j=1}^{2} \partial_{j} \{ A_{j}(\bm{\alpha}) \}
	+ \frac{1}{2} \sum_{j=1}^2 \sum_{k=1}^2 \partial_{j}\partial_{k} \{ D_{jk}(\bm{\alpha}) \} \Big]
	\label{eq:qfpe_apdx}
\end{align}
where $\partial_1 = \partial / \partial \alpha$ and $\partial_2 = \partial / \partial \bar{\alpha}$.
The drift vector $\bm{A}(\bm{\alpha}) = \left( A_1(\bm{\alpha}), A_2(\bm{\alpha}) \right)^{\sf T} \in {\mathbb C}^2$ and the matrix $\bm{D}(\bm{\alpha}) = \left( D_{jk}(\bm{\alpha}) \right) \in {\mathbb C}^{2 \times 2}$ are given by
\begin{align}
	\label{eq:uni_drift}
	\hspace{-3em}
	\bm{A}(\bm{\alpha}) 
	&=
	\left( \begin{matrix}
		\left( \frac{\gamma_1 - \gamma_3}{2} + i \Delta \right) \alpha - \gamma_{2} \overline{\alpha} \alpha^{2} - 2 \eta e^{i \theta}\overline{\alpha}
		\\
		\left( \frac{\gamma_1 - \gamma_3}{2} - i \Delta \right) \overline{\alpha}   - \gamma_{2} \alpha \overline{\alpha}^{2} - 2 \eta e^{ -i \theta}\overline{\alpha}
		\\
	\end{matrix} \right),
	\\
	\label{eq:uni_diffusion}
	\bm{D}(\bm{\alpha}) &= 
	\left( \begin{matrix}
		-( \gamma_{2}  \alpha^{2} + 2 \eta e^{i \theta} )   & \gamma_1  \\
		\gamma_1 & -( \gamma_{2}  \overline{\alpha}^{2} + 2 \eta e^{-i \theta} ) \\
	\end{matrix} \right).
\end{align}
The corresponding SDE of the system, representing the deterministic trajectory subjected to the small quantum noise in the phase space, is given by
\begin{align}
	\label{eq:uni_ldv}
	d
	\left( \begin{matrix}
		{\alpha}  \\
		\overline{\alpha}  \\
	\end{matrix} \right)
	&=
	\left( \begin{matrix}
		\left( \frac{\gamma_1 - \gamma_3}{2} + i \Delta \right) \alpha - \gamma_{2} \overline{\alpha} \alpha^{2} - 2 \eta e^{i \theta}\overline{\alpha}
		\\
		\left( \frac{\gamma_1 - \gamma_3}{2} - i \Delta \right) \overline{\alpha}   - \gamma_{2} \alpha \overline{\alpha}^{2} - 2 \eta e^{ -i \theta}\overline{\alpha}
		\\
	\end{matrix} \right) dt
	+\bm{\beta}(\bm{\alpha})
	\left( \begin{matrix}
		dW_1
		\\
		dW_2
		\\
	\end{matrix}
	\right)
	,
\end{align}
where $W_1$ and $W_2$ are independent Wiener processes,
and the matrix ${\bm \beta}({\bm \alpha})$ is written as
\begin{align}
	\label{eq:uni_beta}
	\bm{\beta}(\bm{\alpha}) &= 
	\begin{pmatrix}
		\sqrt{\frac{\left( \gamma_1 +  R_{11}(\bm{\alpha}) \right)}{2}} e^{i \chi_{11}(\bm{\alpha}) / 2}
		&
		-i \sqrt{\frac{\left( \gamma_1 -  R_{11}(\bm{\alpha}) \right)}{2}} e^{i \chi_{11}(\bm{\alpha}) / 2}
		\\
		\sqrt{\frac{\left( \gamma_1 +  R_{11}(\bm{\alpha}) \right)}{2}} e^{- i \chi_{11}(\bm{\alpha}) / 2}
		&
		i \sqrt{\frac{\left( \gamma_1 -  R_{11}(\bm{\alpha}) \right)}{2}} e^{- i \chi_{11}(\bm{\alpha}) / 2}
	\end{pmatrix},
\end{align}
where $R_{11}(\bm{\alpha}) e^{i \chi_{11}(\bm{\alpha})} = -( \gamma_{2}  \alpha^{2} + 2 \eta e^{i \theta} )$. The two equations for $\alpha$ and $\overline{\alpha}$ in Eq.~(\ref{eq:uni_ldv}) are complex conjugate and describe the same dynamics.

In the classical limit, where the quantum noise vanishes,  the system is characterized by 
a single complex variable $\alpha \in {\mathbb C}$, which obeys the deterministic part of Eq.~(\ref{eq:uni_ldv}), given by  
\begin{align}
	\dot{\alpha} = \left( \frac{ \gamma_1 - \gamma_3}{2} + i \Delta \right) \alpha - \gamma_{2} \overline{\alpha} \alpha^{2} - 2 \eta e^{i \theta} \overline{\alpha}.
	\label{eq:uni_ldv_alpha}
\end{align}

Using the modulus $R$ and argument $\phi$ of the complex variable $\alpha = R e^{i \phi}$, the differential equations for these variables are given by
\begin{align}
	\label{eq:uni_ldv_rad}
	\dot{R} &=  \frac{ \gamma_1  - \gamma_3}{2} R-\gamma_{2} R^{3} - 2 \eta R \cos (2 \phi-\theta), 
	\\
	\label{eq:uni_ldv_phi}
	\dot{\phi} &= \Delta + 2 \eta \sin (2 \phi-\theta).
\end{align}

When $\Delta < 2\eta$, the equation for the argument in Eq.~(\ref{eq:uni_ldv_phi}) has two stable fixed points  
$\phi_{ss} = \frac{1}{2}\left( \pi +  \theta + \arcsin  \left( \frac{\Delta}{2 \eta} \right) \right)$ and $\phi_{ss} + \pi$.
Additionally, when $\gamma_1 \geq \gamma_3$, or $\gamma_1 < \gamma_3$ and $\eta > \frac{1}{2} \sqrt{ \frac{ (\gamma_1 - \gamma_3)^2 }{4} + \Delta^2}$,
the equation for the modulus $R$ has a non-zero solution 
$R_{ss} = \sqrt{\frac{1}{\gamma_2} \left(  \frac{\gamma_1 - \gamma_3}{2} + \sqrt{4\eta^2 - \Delta^2} \right)}$,
and consequently, the system in Eq.~(\ref{eq:uni_ldv_alpha}) possesses two stable fixed points
$\alpha_{ss} = \pm R_{ss} e^{i\phi_{ss}}$.

\begin{figure} [htbp]
	\begin{center}
		\includegraphics[width=0.7\hsize,clip]{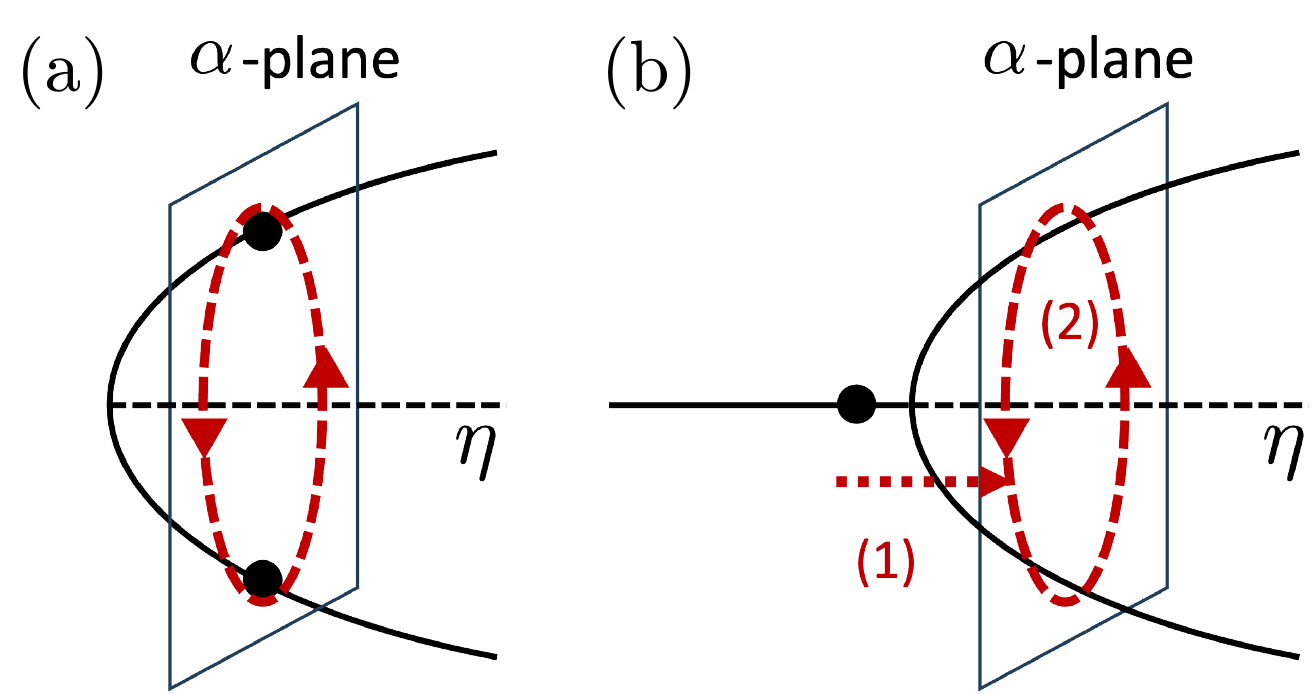}
		\caption{
			A schematic diagram of noise-induced oscillations near the (a) SNIC bifurcation and (b) pitchfork bifurcation  in the classical limit. In (b), the system exceeds the bifurcation points of the (1)
			modulus and (2) argument simultaneously.}
		\label{fig_6}
	\end{center}
\end{figure}

Figs.~\ref{fig_6} (a) and~\ref{fig_6}(b) schematically depict the bifurcation diagram when varying $\eta$ for cases in Sec. \ref{sec:qvdp_sq} and Sec. \ref{sec:dpo}, respectively.  In Sec. \ref{sec:qvdp_sq}, the equation for the argument  in Eq.(\ref{eq:uni_ldv_phi}) is slightly below the bifurcation, i.e., $\Delta < 2\eta$, and the equation for the modulus in Eq.(\ref{eq:uni_ldv_rad}) always has two stable fixed points as $\gamma_1 >  \gamma_3 (= 0)$.
Then, as illustrated in Fig.~\ref{fig_6}(a),  the oscillatory response is excited by the quantum noise, characterized by a round trip around the two stable fixed points . 
 In Sec. \ref{sec:dpo}, the equation for the argument in Eq.(\ref{eq:uni_ldv_phi}) is slightly below the bifurcation, i.e., $\Delta < 2\eta$, and with $\gamma_3 >  \gamma_1 (= 0)$, the equation for the modulus in Eq.(\ref{eq:uni_ldv_rad}) is slightly below the bifurcation, i.e., $\eta <  \frac{1}{2} \sqrt{ \frac{ (\gamma_1 - \gamma_3)^2 }{4} + \Delta^2}$. 
 Then, as illustrated in Fig.~\ref{fig_6} (b), due to the quantum noise, the system exceeds the pitchfork bifurcation point of the modulus and two stable fixed points emerge, and simultaneously, the system exceeds the bifurcation point of the argument. As a result, the system exhibits oscillatory response of  round trip around the two emerging stable fixed points. Thus, the system exhibits quantum noise-induced oscillations starting from the origin, arising because the system simultaneously exceeds the bifurcation points of both the modulus and argument.

%%%%%%%%%%%%%%%%%%%%%%%%%%%%%%%%%%%%%%%%%%%%%%%%%%%%%%%%%%%%%%%%%%%%%%%%
%
\section{Amplitude function of a harmonic oscillator with linear damping}
\label{sec:apdx_damp_ho}

In Ref.\cite{perez2021isostables}, the stochastic asymptotic amplitude function for a classical stochastic damped harmonic oscillator was studied. 
In this appendix, building upon their findings, we calculate the quantum asymptotic amplitude for a simple quantum harmonic oscillator with linear damping.
Although the system behaves linearly in the classical limit and does not exhibit nonlinear oscillations, we can still introduce the asymptotic  amplitude function, as defined in Sec. \ref{sec:quantum_qiso}.
In this setting, we can analytically obtain the eigenoperator $V_2$ of the adjoint Liouville operator ${\mathcal L}^*$.
The quantum master equation of a damped harmonic oscillator is given by  
\begin{align}
	\label{eq:damp_ho_me}
	\dot{\rho}
	= {\mathcal L}\rho 
	= -i[ {\omega_0} a^{\dag} a, \rho] 
	+  \gamma_3 \mathcal{D}[a]\rho.
\end{align}
For simplicity, we assume $\omega_0 \neq 0$ and the system is oscillatory. 

The eigenoperator associated with the largest non-zero eigenvalue on the real axis of the adjoint Liouville operator ${\mathcal L}^*$ is explicitly obtained as $V_2 = a^{\dag} a$, i.e., ${\mathcal L}^* a^{\dag} a = \kappa_q a^{\dag} a$ with
$\kappa_q = - \gamma_3$~\cite{barnett2000spectral, briegel1993quantum}.
Therefore, the quantum asymptotic amplitude function $R_q(\bm{\alpha})$ of the coherent {state} $\bm{\alpha}$ is given by
\begin{align}
	R_q(\bm{\alpha})  = \bra{\alpha} a^{\dag} a \ket{\alpha} = \abs {\alpha}^2,   
\end{align}
and the asymptotic amplitude function $R_q(\rho)$ of the density operator $\rho$ is given by
\begin{align}
	R_q(\rho) = \langle a^{\dag} a \rangle_q =  \langle \rho, a^{\dag} a \rangle_{tr} \left(= \int \abs{\alpha}^2 p({\bm \alpha}) d{\bm \alpha} \right).
\end{align}
For the initial condition $\rho_0 = \ket{\alpha_0} \bra{\alpha_0}$
with $\alpha_0 = r_0 e^{i \theta_0}$, the expectation of $a^{\dag} a$ evolves as
\begin{align}
	\frac{d}{dt} \langle a^{\dag} a \rangle_q
	= \langle \dot{\rho}, a^{\dag} a \rangle_{tr} 
	= \langle {\mathcal L} \rho, a^{\dag} a \rangle_{tr} 
	= \langle \rho, {\mathcal L}^* a^{\dag} a \rangle_{tr} 
	= \kappa_q \langle \rho, a^{\dag} a \rangle_{tr} 
	= \kappa_q \langle a^{\dag} a \rangle_q,
\end{align}
which yields $\langle a^{\dag} a \rangle_q = e^{\kappa_q t}  \langle \rho_0, a^{\dag} a \rangle_{tr}   = e^{\kappa_q t} \bra{\alpha_0} a^{\dag} a \ket{\alpha_0} = e^{\kappa_q t} \abs{\alpha_0}^2 = e^{ -\gamma_3 t}r_0^2
$.

Thus, the asymptotic amplitude of the state $\rho$ is obtained as
\begin{align}
	R_q(\rho) =  e^{ -\gamma_3 t} r_0^2,
\end{align}
which decays exponentially to zero with a constant rate $\gamma_3$. 
Consequently, the zero level-set of the asymptotic amplitude forms a stable fixed point at the origin,  and the effective quantum periodic orbit $\chi_q$ is absent in this case. 

As in the example of a quantum harmonic oscillator with linear damping, the asymptotic amplitude function can be formally introduced  for a wide variety of systems, even if the system lacks nonlinear oscillatory dynamics. 
%

%\bibliographystyle{apsrev4-1}
%\bibliographystyle{unsrt}
%\bibliography{reference_row}

\end{document}